# Metal-insulator transition in vanadium dioxide nanobeams: probing sub-domain properties of strongly correlated materials

*Jiang Wei†, Zenghui Wang†, Wei Chen†, and David H. Cobden†**
† Department of Physics, University of Washington, Seattle WA 98195-1560

* corresponding author, email cobden@u.washington.edu

**Many strongly correlated electronic materials, including high-temperature superconductors, colossal magnetoresistance and metal-insulator-transition (MIT) materials, are inhomogeneous on a microscopic scale as a result of domain structure or compositional variations. An important potential advantage of nanoscale samples is that they exhibit the homogeneous properties, which can differ greatly from those of the bulk. We demonstrate this principle using vanadium dioxide, which has domain structure associated with its dramatic MIT[1-3] at 68 ˚C. Our studies of single-domain $VO_2$ nanobeams reveal new aspects of this famous MIT, including supercooling of the metallic phase by 50 ˚C; an activation energy in the insulating phase consistent with the optical gap; and a connection between the transition and the equilibrium carrier density in the insulating phase. Our devices also provide a nanomechanical method of determining the transition temperature, enable measurements on individual metal-insulator interphase walls, and allow general investigations of a phase transition in quasi-one-dimensional geometry.**

At temperatures above $T_c$ = 68 ˚C, bulk $VO_2$ is a poor metal, while below $T_c$ it is a semiconductor with an optical gap[4] $E_g$ = 0.6 eV. The transition to the metal can be induced very rapidly and has recently been studied intensively by ultrafast techniques.[5-11] The lattice in the metallic phase has the rutile structure, with the vanadium ions arranged in periodic chains parallel to the *c*-axis. In the insulating phase these are distorted into dimerized zig-zag chains, resulting in a monoclinic structure known as M1. Many factors indicate that the transition involves strong electron-electron correlations, as in a Mott transition. These include the anomalously low conductivity and other properties of the metal;[12-14] the fact that band structure calculations fail to yield the insulator band gap;[3,15] the fact that an intermediate M2 structure, which can be stabilized by stress[16] or doping,[17] is insulating in spite of having undimerized vanadium chains;[18] and in optical experiments a dependence on excitation power which indicates sensitivity to excited carrier density.[9,10,19] However, the nature of the transition is still quite unclear half a century after its discovery,[1] and potential applications for instance in electrical[20] and optical[21] switching or detection remain unrealized. The blame for this falls largely on the domain structure produced on passing through the MIT which leads to irreproducibility between samples (properties such as resistivity are very sensitive to the arrangement of domains), broadening and hysteresis of the characteristics, and nonuniform stresses producing mechanical degradation.[12] All these problems are absent in our nanobeam devices, which are smaller than the characteristic domain size.

The nanobeams were grown directly on oxidized Si wafers by a physical vapor transport technique based on that introduced by Park's group:[22,23] the substrate is placed in 20 mbar Ar carrier gas downstream from a granular $VO_2$ source in a tube furnace at 1000 °C for 30 minutes. Each nanobeam is a single crystal elongated along the rutile *c*-axis. The length can be hundreds of microns and the cross-section is roughly rectangular, the width $W$ ranging down to 50 nm and thickness $H$ down to 15 nm as determined by atomic force microscopy. As reported by Wu *et* *al.*,[23] on warming, quasiperiodic thin stripes of darker metallic phase can be seen to appear around 60 °C in an optical microscope (Fig. 1a). These stripes then widen and eventually merge, and the insulating phase has disappeared by around 105 °C. This behaviour results from strain caused by firm attachment to the $SiO_2$ substrate. A fully insulating nanobeam is under compressive axial strain, while a fully metallic one, having a smaller equilibrium *c*-axis lattice constant, is under tension. Alternation of metallic and insulating regions reduces the average strain at the cost of creating interphase walls. Wu *et al.* also found that if the nanobeams are released from the substrate, completely relieving the strain, the transition becomes sudden with no stable domain pattern formation.

Here, rather than eliminating the axial strain we exploit it. Using electron-beam or optical lithography we pattern a series of electron-beam evaporated metal contacts (typically 10 nm vanadium under 400 nm gold) onto a nanobeam, and then immerse in buffered oxide etch to remove the 1 μm-thick $SiO_2$ beneath it where it is not covered by metal and thereby to suspend the nanobeam sections between the contacts. The behaviour of the resulting devices, under repeated cycling in air between 120 °C and room temperature, $T_{room}$, can be reproducible over months, although it is modified significantly by passing large currents or storage in vacuum. At $T_{room}$, shorter sections are straight while longer ones are buckled (see Fig. 1b). This is consistent with the behaviour of clamped beams, in which Euler buckling occurs when the compressive axial stress $P$ exceeds a critical value $P_b$ that is smaller for longer beams. Firm adhesion to the substrate under the contacts provides the clamping. After buckling the compressive strain $\eta$ is much smaller and the nanobeam adopts approximately its natural length $L_0$. By taking $L_0$ to be the length along the curved profile of a buckled section measured in an atomic force microscope, and calling the contact separation $L$, we find that the substrate-induced strain in most nanobeams at $T_{room}$ is $\eta_0 = (L_0-L)/L = 0.4\pm0.1$ %.

Seen in an optical microscope, between about 68 and 105 °C every segment of each nanobeam is unbuckled and consists of one metallic and one insulating region coexisting and separated by a single interphase wall (see Fig. 1c). As $T$ increases, the fraction $x$ of insulating phase, plotted in Fig. 1d, decreases steadily until at a temperature $T_m$ in the vicinity of 105 °C the nanobeam becomes fully metallic. This behaviour can be understood by noting that given fixed $L$, once the nanobeam begins to convert from insulator to metal its equilibrium length $L_0$ decreases because of the shorter *c*-axis of the metal, and so therefore do both $\eta$ and $P$. At a particular $T$ the fraction $x$ of insulator adjusts so that $P$ is appropriate for the two phases to coexist, ie, to lie on the phase boundary line between insulator and metal in the $(P,T)$-plane, as sketched in Fig. 1e. The stress $P$ should be zero at $T = T_c$ and negative (tensile) at higher $T$, explaining why all nanobeams become straight above about 68 °C.

The nanobeam thus provides a one-dimensional (1D) analog of the 3D situation of water held at constant volume near 0 ˚C: there too the low-$T$ phase (ice) has higher volume and lower symmetry than the high-$T$ phase (liquid water), and the fraction $x$ of ice decreases as $T$ increases in a corresponding way. It is also worth noting that in the coexistence regime the nanobeam has zero axial stiffness, just as the ice-water mixture has diverging compressibility.

If we assume the Young's modulus $E$ is the same for both phases then the strain $\eta(T)$ is uniform along the nanobeam and the equilibrium phase boundary line $P(T)$ is determined by

$$P(T)/E = \eta(T) = \alpha\,(x - x_c) + K\,(T - T_c)\,. \qquad (1)$$

The first term on the right represents interconversion between the phases, where $\alpha$ is the fractional increase in rutile c-axis length going from metal to insulator and $x_c$ is the insulating fraction at $T = T_c$. The second term represents differential thermal expansion of the $VO_2$ relative to the Si substrate. It is an order of magnitude smaller than the first term, $K$ being about $+2.0\times10^{-5}$ °C$^{-1}$ and assumed the same for both phases.[23-25]

According to equation (1), for every nanobeam $x$ should have the same variation with $T$ to within an offset $x_c$ which depends on the built-in strain relative to the substrate and may vary with growth conditions. The data in Fig. 1d agree well with this prediction, supporting the assumptions made above. Moreover we observe that $x(T)$ is nearly a straight line, with $dx/dT = -(1.10\pm0.05)\times10^{-2}$ °C$^{-1}$. This implies that the phase boundary line is nearly straight over the experimental temperature range so that we can write simply

$$P(T)/E = \beta\,(T - T_c)\ , \qquad (2)$$

where $\beta = \alpha\, dx/dT+K$ is a constant. According to equation (2), $P$ will become positive when a nanobeam in coexistence is cooled below $T_c$, and upon further cooling we expect the nanobeam to buckle at the temperature $T_b$ for which $P(T_b) = P_b$. Using the Euler expression for the buckling pressure of a doubly clamped beam, $P_b = (\pi^2E/3)H^2/L^2$, we then have

$$\beta\,(T_b - T_c) = (\pi^2/3)H^2/L^2\ . \qquad (3)$$

Equation (3) predicts that a plot of $T_b$ vs $1/L^2$ will yield a straight line with $y$-intercept $T_c$. We make such a plot in Fig. 1f for a series of sections of a nanobeam of thickness $H$ = 180±5 nm. The data are indeed well fitted by the straight line shown, with $y$-intercept $T_c$ = 65.7±0.2 °C.

This represents a completely new way of measuring the transition temperature, exploiting coupling of the phase transition to nanomechanical motion. Unlike other methods it is independent of hysteresis at the transition. In fact, the more general procedure of finding $\lim_{L\to\infty}(T_b)$ should yield $T_c$ independently of the assumptions made above, provided only that the phase boundary is well behaved. This could be exploited for example to methodically study variations in the phase boundary between samples or in response to modified external conditions.

We can also use equation (3) to obtain $\beta$ from the slope $dT_b/d(L^{-2}) = -820\pm20$ °C μm$^2$ of the line, giving $\beta = (\pi^2H^2/3)/[\partial T_b/\partial(L^{-2})] = (-13 \pm 1)\times10^{-5}$ °C$^{-1}$. Using the estimate[26] $E \approx 140$ GPa, we can then quantify the stress in the nanobeam: the phase boundary slope is $\beta E \approx -18$ MPa °C$^{-1}$, and the largest tension, reached at $T = T_m \approx 110$ °C, is $P_m = \beta E\,(T_m - T_c) \approx -0.7$ GPa. For $T > T_m$ the nanobeam is fully metallic and thermal expansion causes the tension to decrease again, at a rate $dP/dT = KE \approx +2.5$ MPa °C$^{-1}$, as indicated in Fig. 1e. Also, using this value for $\beta$ we find $\alpha = (\beta - K)/(dx/dT) = 1.36 \pm 0.15$ %. The discrepancy with the accepted value[3] of $\alpha = 1.0$ % could be partly due to the phases actually having different elastic moduli, as well as to another factor which we comment on at the end.

Measurements of the two-terminal electrical resistance $R$ in the coexistence regime yield a number of interesting results. Fig. 2a shows the characteristics on repeated thermal cycling of four nominally equal sections (A–D) of a single nanobeam. Each comprises smooth $R$ vs $T$ curves punctuated by sudden jumps. On warming from $T_{room}$, $R$ initially decreases in a semiconducting manner. It drops sharply at a temperature $T_n \sim 65$–68 °C when a metallic region appears and the nanobeam becomes straight. Subsequently, in the coexistence regime, as the insulating region shrinks $R$ decreases steadily to a much smaller value at $T_m \sim 105$ °C at which the insulator disappears. The coexistence curve is reproducible on sweeping up and down at 6 °C/min as long as $T$ is kept below $T_m$. On the other hand, on cooling from above $T_m$ each section remains fully metallic down to a lower temperature $T_s$ at which an insulating region suddenly appears and $R$ jumps up to the coexistence curve. $T_s$ varies greatly between sections, from 78 ˚C (section A) to 55 ˚C (section D). During further cooling, $R$ increases smoothly until the nanobeam buckles at $T_b$ = 57.5 °C. Below that, the temperature $T_i$ at which $R$ returns to the insulating curve also varies, as several conformations are possible after the nanobeam initially buckles downwards and collides with the substrate. When sweeping at 6 °C/min, $T_s$ and $T_n$ vary within a range of 2–5 °C between sweeps, while $T_b$ and $T_m$ are reproducible to within the measurement accuracy of about 0.1 ˚C.

The supercooling of the homogeneous, uniformly stressed metallic phase to $T_s$, which can be more than 50 °C below the phase boundary, is represented by the green lines on the phase diagram in Fig. 2b. This intrinsic supercooling (as opposed to the persistence of metallic domains due to inhomogeneous strain fields in larger samples) is much larger than has been reported previously[23] and indicates high crystal uniformity. The large variations in $T_s$ between nanobeams could reflect the availability of imperfections at which the insulating phase can nucleate. This system offers new opportunities for investigating the kinetics of a first-order phase transition in a quasi-1D geometry, for example by studying the process of insulator nucleation at $T_s$ or the disappearance of the insulating domain at $T_m$ (see inset to Fig. 2a).

In Fig. 3 we plot the resistivity $\rho_i$ (along the rutile $c$-axis) of a number of nanobeams in their fully insulating state obtained simply using $\rho = RA/L$, where $A = WH$. Studies of the dependence on $L$ showed that contact resistance was negligible. $\rho_i$ is consistent between samples and its $T$ dependence gives an activation energy $E_a$ = 0.30±0.01 eV, indicated by the dashed line. Interestingly, this activation energy corresponds to that expected for an intrinsic semiconductor with precisely the established optical gap[4] $E_g$ = 0.60 eV of $VO_2$. This is in contrast with the anomalously large and sample-dependent activation energies of up to 0.45 eV reported in the literature[12] on bulk insulating $VO_2$ which are most likely influenced by domain structure or imperfections that are absent in the nanobeams.

We also plot the resistivity $\rho_m$ of the metallic phase for one nanobeam. Consistent with the literature[13] we find that $\rho_m$ increases slowly with $T$ and that at the transition $\rho_i \approx 2\times10^4\rho_m$. The kink in $\rho_m$ at 107 ˚C, indicated by an arrow, occurs at the boundary between supercooled and stable metallic phase. We do not know its origin, although we note that recent experiments have indicated that structural transitions may occur within the metallic phase under hydrostatic pressure[27] or transiently after ultrafast excitation.[7]

In the coexistence regime we expect $R$ to have four contributions:

$$R = x\rho_i^c L/A + (1\text{-}x)\rho_m^c L/A + R_c + R_{dw}\,, \qquad (3)$$

where $\rho_i^c$ and $\rho_m^c$ are the insulator and metal resistivities in coexistence, and $R_c$ and $R_{dw}$ are the resistances of the contact to the insulator and of the domain wall respectively. Since $\rho_i \gg \rho_m$ the second term is negligible. At $T_m$ when the insulating domain disappears, there is usually a small sudden drop in $R$ of around 50 kΩ (see inset to Fig. 2a). We have yet to determine whether this reflects the resistance of the domain wall, the minimum insulating domain size, or the contact to the insulator, but in any case it implies that $R_c + R_{dw} \lesssim 50$ kΩ. Since $R$ is usually greater than 1 MΩ for measurable $x$, to good accuracy we can thus write

$$R = \rho_i^c\, xL/A\,. \qquad (4)$$

As a result, the resistance can act as a probe of the position $xL$ of the interphase wall. We find that $R$ can be stable to within 0.1 %, limited by the thermal stage stability; hence it can be used to detect motion of the wall with an accuracy of less than 10 nm in a device with $L$ = 10 μm, offering the possibility to detect small effects of perturbations such as electric field on the transition. Applying this idea, we found no change in $R$ on applying up to 50 V to the Si substrate, indicating that the screening length within the semiconductor is less than the minimum nanobeam thickness, ~30 nm, and implying that Mott transistor action[20] will be difficult to achieve.

Most interestingly of all, by equation (4) we can determine $\rho_i^c$ from $R$ using values of $x$ obtained by optical inspection. The results are included in Fig. 3. We find that $\rho_i^c = 12\pm2\ \Omega$ cm for all the nanobeams, independent of $T$ to within measurement error. This is in sharp contrast with the activated behavior of $\rho_i$ which is surely due to carrier activation across an unchanging gap. It implies that the carrier density in the insulator becomes independent of $T$ once the nanobeam enters coexistence; that is, in coexistence the changing strain causes the gap to increase in just such a way as to counteract the effect of thermal activation on the carrier density. In other words, the phase boundary corresponds to a contour of constant insulator carrier density in the $P$-$T$ plane.

This remarkable result is unlikely to be a coincidence, especially considering recent evidence that the excited carrier density in the insulator is important when the transition is induced optically.[9,10,19] Rather, it implies that the transition temperature is closely linked to the equilibrium carrier density in the insulating phase. This is consistent with a driving role for electron-electron interactions in the MIT, as in a Mott transition which occurs when the density-dependent screening reaches a critical strength. In contrast, a phonon-driven mechanism would not be expected to be sensitive to the nondegenerate carrier density in the insulator.

These new results on the MIT have been made possible by the greatly improved reproducibility, uniformity and control achievable in single-domain samples compared with the already intensively studied bulk samples. Amongst other aspects of the system that we are still investigating, one should be mentioned here: the $VO_2$ nanobeams usually pass through an intermediate insulating state, most probably the M2 phase, during the transition between M1 and metallic phases (this explains why we found a larger value of $\alpha$ than expected). We conclude by reiterating that this work demonstrates that there are many significant insights and advantages to be gained by studying strongly correlated materials in nanoscale crystalline form.

**Acknowledgments**

We thank Anton Andreev, Boris Spivak. Oscar Vilches and Younan Xia for discussions, Volker Eyert and Hyun-tak Kim for comments on the manuscript, and Jacob Beedle, Megan Campbell and Conor Sayres for experimental assistance. This work was supported by the Army Research Office under contract number 48385-PH, and employed facilities in the UW Nanotech Center. WC and ZW were partially supported by UW UIF Nanotech fellowships.

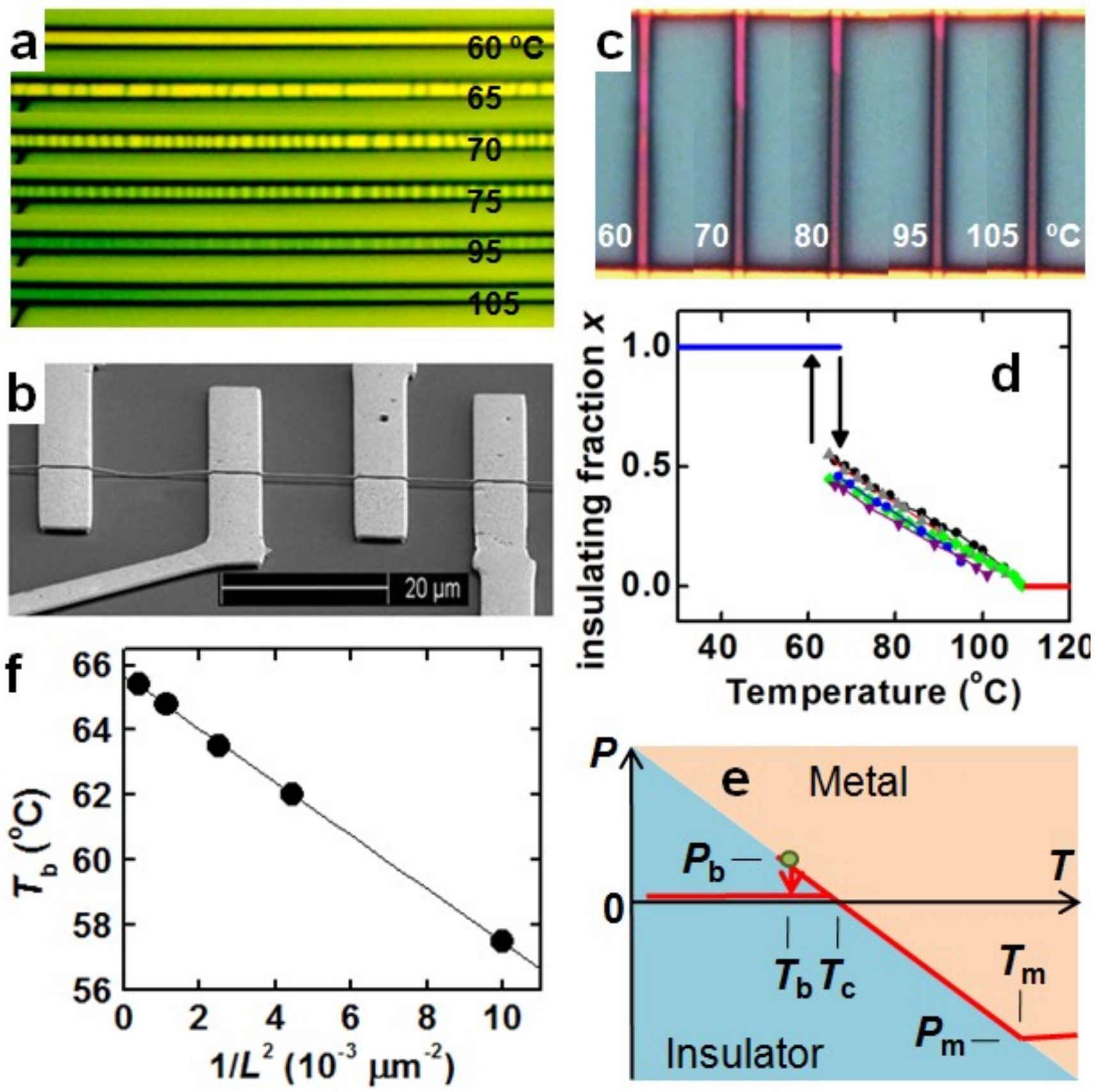

**Figure 1** Metal-insulator transition in $VO_2$ nanobeams studied by microscopy. **a**, Six images of a 40 μm-long part of a single nanobeam attached as grown to an $SiO_2$ substrate, taken at the indicated temperatures during warming, showing metallic domains (darker) appearing, widening and merging. **b**, SEM image of a suspended nanobeam device showing that longer sections are buckled at room temperature. **c**, Five images of one suspended nanobeam between contacts at the top and bottom separated by 20 μm. Above 68 °C it contains a single metallic domain (grey), which grows on warming until the insulating domain (purple) disappears at about 105 ˚C. **d**, Plot of the insulating fraction *x* vs. *T* for six suspended nanobeams of various dimensions. **e**, Sketch of the phase diagram indicating part of the trajectory (red) followed on a temperature cycle. The vertical axis is uniaxial pressure *P*. The green circle indicates the point at which the nanobeam buckles on cooling. **f**, Plot of buckling temperature $T_b$ vs. inverse square length for a nanobeam (thickness $H = 0.18$ μm, width $W = 0.9$ μm) yields a straight line whose *y*-intercept should be $T_c$.

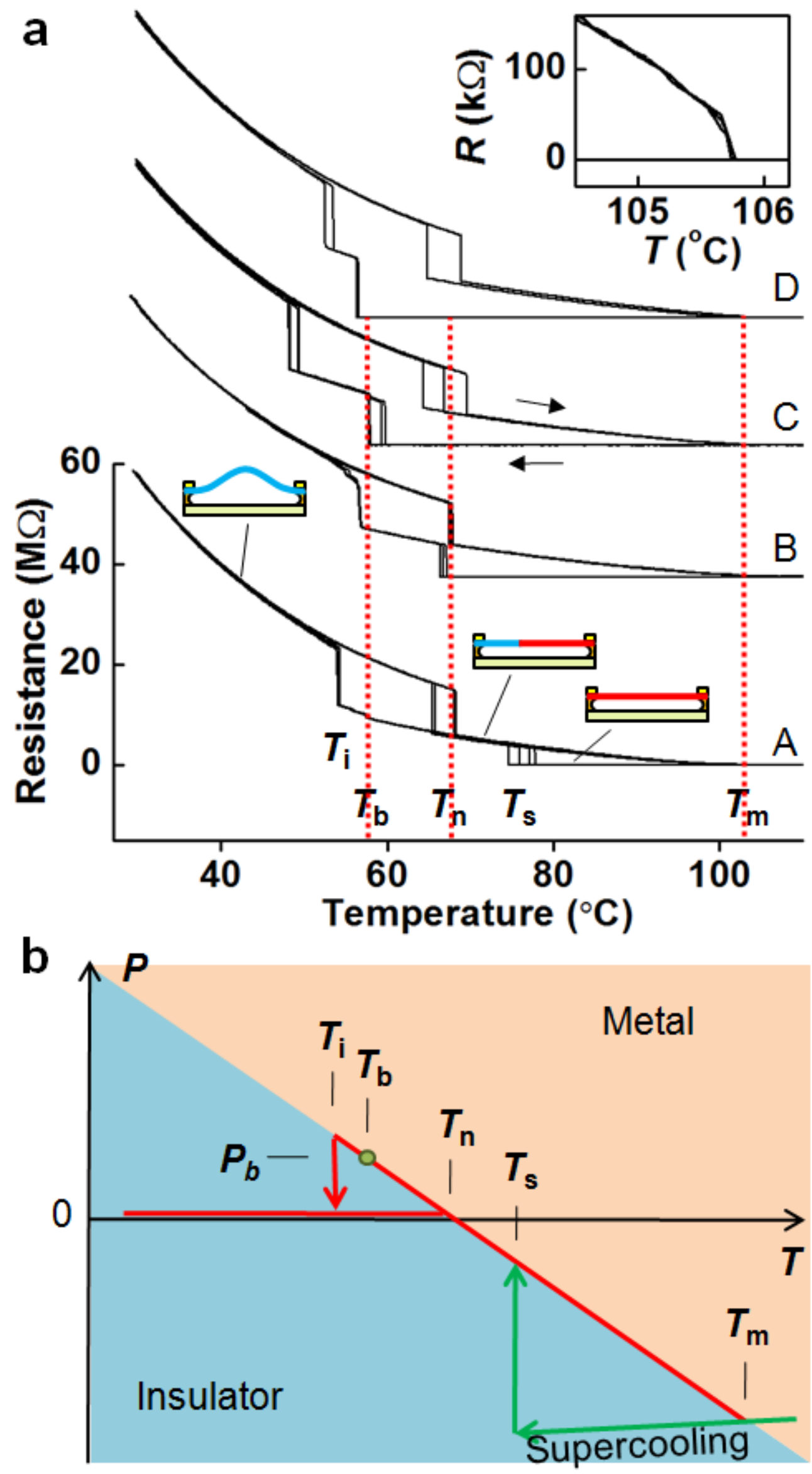

**Figure 2** Electrical resistance vs temperature. **a**, Characteristics of four sections, designated A-D, of the same nanobeam, each having $L = 26$, $H = 0.25$, and $W = 0.7$ µm. The three stable states of each section are indicated by the sketches: buckled and fully insulating (blue), straight and fully metallic (red), or coexisting (part blue, part red). Upper inset: a repeatable small jump of about 50 kΩ is seen as the insulator disappears near $T_m$. **b**, Corresponding phase diagram, where $P$ is the axial pressure. The red line indicates the trajectory followed by section A. On warming the nanobeam unbuckles and jumps from fully insulating to coexistence at $T_n$, and becomes fully metallic at $T_m$. On cooling it supercools to $T_s$ at which the insulator nucleates, it buckles downwards at $T_b$, and it then buckles upwards and becomes fully insulating at $T_i$.

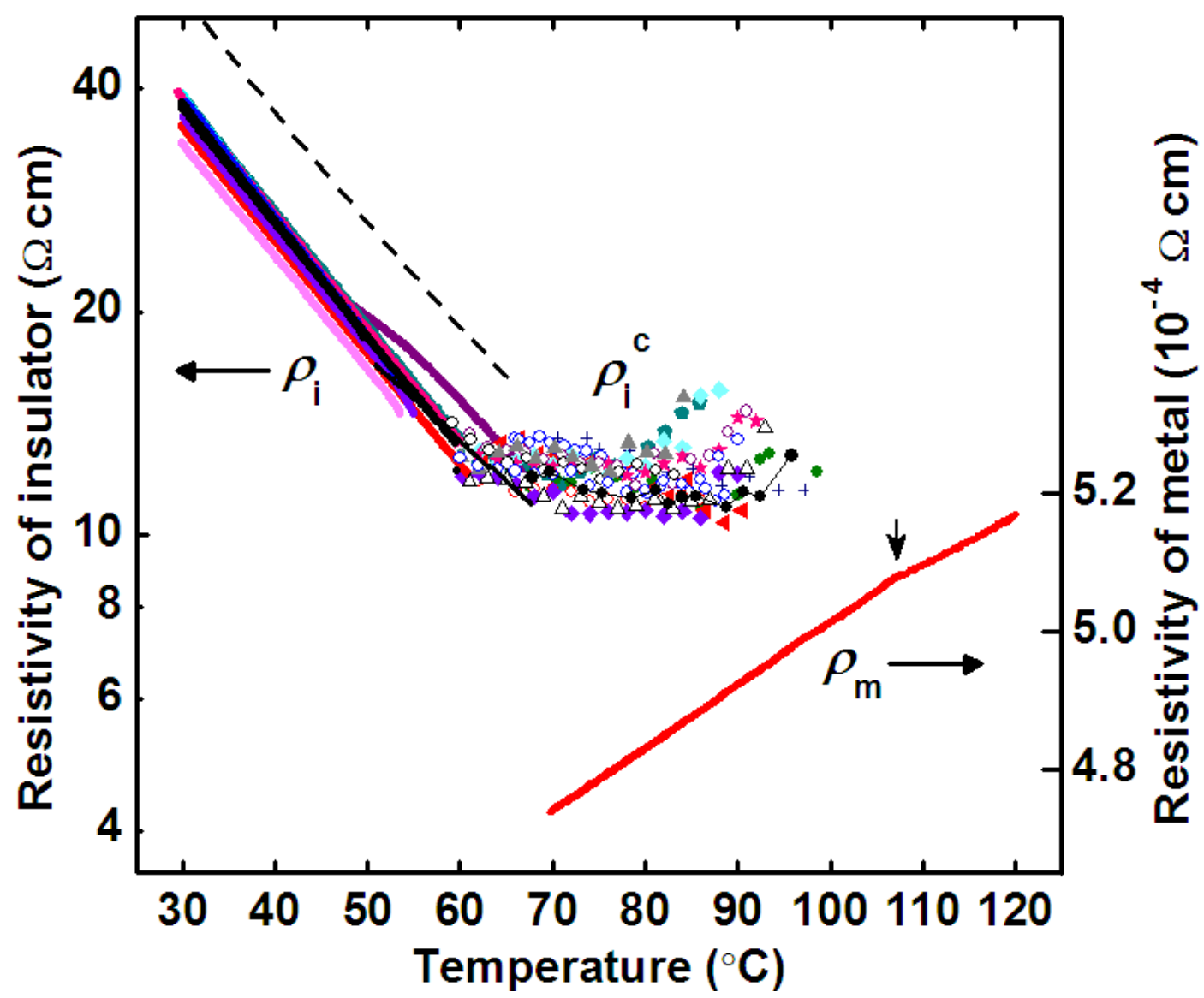

**Figure 3** Collected resistivity measurements for ten nanobeams of various dimensions. The insulator resistivity $\rho_i$ exhibits an activation energy $E_a$ of 0.30 eV (dashed line is $e^{-E_a/kT}$). Plotted using symbols are measurements of the resistivity $\rho_i^c$ of the insulator in coexistence for ten different nanobeams. To within error they all show a temperature-independent value of 12 ± 2 Ωcm. The metal resistivity $\rho_m$ was obtained from one section of a nanobeam ($H = 0.18$, $W = 0.9$, $L = 30$ μm). Two other sections of the same nanobeam (with $L = 20$ and 50 μm) gave almost identical results. We estimate the error in $\rho_m$ due to uncertainty in the cross-section to be 25%. The values of $\rho_m$ for temperatures below the kink (small arrow) are in the supercooled regime.